\documentclass[aps,nofootinbib,prd,aps,epsf,floats,floatfix,amsmath,amssymb,amsfonts,axodraw,floatfix,graphicx,twocolumn]{revtex4-2}  
\usepackage{graphicx}  
\usepackage{dcolumn}   
\usepackage{bm}        
\usepackage{amssymb}   
\usepackage{amsmath}
\usepackage{etoolbox}
\usepackage{enumitem}
\usepackage{stackrel}
\usepackage{float}
\usepackage[normalem]{ulem}
\usepackage{mathtools}
\renewcommand{\theequation}{\thesection.\arabic{equation}}
\newcounter{saveeqn}
\newcommand{\add}{\addtocounter{equation}{1}}
\newcommand{\alpheqn}{\setcounter{saveeqn}{\value{equation}}%
	\setcounter{equation}{0}%
	\renewcommand{\theequation}{\mbox{\thesection.\arabic{saveeqn}{\alph{equation}}}}}
\newcommand{\reseteqn}{\setcounter{equation}{\value{saveeqn}}%
	\renewcommand{\theequation}{\thesection.\arabic{equation}}}

\newsavebox{\notrightarrow}
\sbox{\notrightarrow}{$\to$\hspace{-4mm}/}

\newsavebox{\PARTIALSLASH}
\sbox{\PARTIALSLASH}{$\partial$\hspace{-1.6mm}/}

\newsavebox{\ASLASH}
\sbox{\ASLASH}{$A$\hspace{-2.1mm}/}

\newsavebox{\KSLASH}
\sbox{\KSLASH}{$k$\hspace{-1.8mm}/}

\newsavebox{\LSLASH}
\sbox{\LSLASH}{$\ell$\hspace{-1.8mm}/}

\newsavebox{\QSLASH}
\sbox{\QSLASH}{$q$\hspace{-1.8mm}/}

\newsavebox{\DSLASH}
\sbox{\DSLASH}{$D$\hspace{-2.2mm}/}

\newsavebox{\DbfSLASH}
\sbox{\DbfSLASH}{${\mathbf D}$\hspace{-2.8mm}/}

\newsavebox{\DELVECRIGHT}
\sbox{\DELVECRIGHT}{$\stackrel{\rightarrow}{\partial}$}

\newcommand{\blue}{\IfColor{\textCadetBlue}{}}
\newcommand{\black}{\IfColor{\textBlack}{}}
\newcommand{\red}{\IfColor{\textRed}{}}
\newcommand{\green}{\IfColor{\textOliveGreen}{}}
\newcommand{\lila}{\IfColor{\textRedViolet}{}}

\newcommand{\MeV}{\mathop{\rm MeV}\nolimits}

\newcommand{\fmc}{\mathop{\rm fm/c}\nolimits}

\newcommand{\rd}[1]{\mathop{\mathrm{d}#1}}

\newcommand{\deriv}[2]{\frac{{d}#1}{{d}#2}}

\newcommand{\inv}[1]{\frac{1}{#1}}

\newcommand{\dform}[1]{\mathbf{#1}}
\newcommand{\cd}[2]{\mathrm{\nabla}_{#1}{#2}}

\newcommand{\lieder}[2]{\mathcal{L}_{#1}{#2}}

\newcommand*{\derivn}[3][]{\ensuremath{\frac{{d}^{#1} #2}{{d} #3^{#1}}}}

\DeclareRobustCommand{\orderof}{\ensuremath{\mathcal{O}}}

\DeclareMathOperator{\dext}{\mathbf{d}}

\DeclareBoldMathCommand{\xbot}{{\bf x}_\perp}
\newcommand{\tvec}[1]{\boldsymbol{#1}}

\renewcommand\u{\underline}
\newcommand\NFSYM{\mathcal{N}=4~\text{SYM}}

\DeclareRobustCommand{\sexp}{\ensuremath{\nabla\cdot u}}
\DeclareRobustCommand{\ceta}{\ensuremath{C_\eta}}
\DeclareRobustCommand{\cpi}{\ensuremath{C_p}}
\DeclareRobustCommand{\ctheta}{\ensuremath{C_Q}}
\DeclareRobustCommand{\ed}{\ensuremath{\epsilon}}
\DeclareRobustCommand{\tempr}{\ensuremath{\mathcal{T}}}

\begin{document}



\title{Conformal Bjorken flow in the general frame and its attractor\\ Similarities and discrepancies with the M\"uller-Israel-Stewart formalism}
\author{M.~Shokri}
\email{mshokri@ipm.ir}
\affiliation{IPM, School of Particles and Accelerators,
	P.O. Box 19395-5531, Tehran, Iran}
\author{F.~Taghinavaz}%
\email{ftaghinavaz@ipm.ir}

\affiliation{IPM, School of Particles and Accelerators,
	P.O. Box 19395-5531, Tehran, Iran}

\date{\today}

\begin{abstract}
We investigate the implications of the general frame approach for conformal Bjorken flow beyond the earlier studies. We show that the power series solution at late times is not unique and is accompanied by an exact solution of the form $1/\tau$, which becomes unphysical if taken on shell. In contrast to the M\"uller-Israel-Stewart formalism, a matching between $\NFSYM$ results and the hydro expansion is only possible up to the first order, which gives rise to $\eta/s=1/4\pi$. Matching the results to the next order gives rise to causality/stability-violating values.  Furthermore, we show that the pressure anisotropy in the general frame cannot capture the hydrodynamization, and we introduce an alternative measure to find the attractor. Using slow-roll expansion, we find an analytical approximation form for the attractor. We also show that the early-time behavior of attractors is related to stability and causality conditions. The attractor solutions outside the stable and causal regime give rise to reheating and negative longitudinal pressures in early times, in contrast to the stable and causal ones. We also comment on the violation of the second law of thermodynamics by the off-shell parameters. We show that for the stable and causal choice of parameters, the off-shell canonical entropy of the attractors, which is not a physical quantity, has a negative divergence in early times before tending to its on-shell limit. On the other hand, the unstable and acausal attractors have non-negative entropy divergence. We speculate that the violation of the second law by stable and causal off-shell parameters is required for stability of the first-order hydrodynamics. We investigate the analytical structure of the Borel-transformed series and find the proper relation between the poles and nonhydro modes.
\end{abstract}

\maketitle
\section{Introduction}\label{sec:intro}
\setcounter{equation}{0}
\par	
Relativistic hydrodynamics (RH) offers an effective approach to understanding different many-body systems and in particular, the quark-gluon plasma (QGP) produced in heavy-ion collisions (HICs)~\cite{A1-Romatschke, A2-Romatschke, A3-Rezzolla, A4-Kovtun}. The word ``effective'' here means that the infinite microscopic degrees of freedom (d.o.f's) are replaced by a few macroscopic variables such as energy density, pressure, and fluid velocity~\cite{A4-Kovtun}. 
One can regard RH as an extended version of relativistic thermodynamics in which thermodynamic quantities acquire spacetime dependency. It is built upon two assumptions: the existence of stable local thermodynamic states in equilibrium and the possibility of a gradient expansion around such states out of the equilibrium. Therefore in equilibrium, the conserved charges are written in terms of a chosen set of hydrodynamic variables that completely determine the local thermal state. In the zeroth order, no derivative term appears and the resulting conservation laws describe the evolution of a perfect fluid. To incorporate the dissipative effects one adds derivatives of the hydrodynamic variables up to an arbitrary order of truncation~\cite{A4-Kovtun}. If the fluid is not far from the (local) equilibrium one anticipates that truncation of the gradient expansion at first order might be a good approximation. 
\par
A set of hydrodynamic variables that are commonly used in RH includes the temperature, fluid velocity, and chemical potential. Although these quantities are well defined in equilibrium, they lose their uniqueness when the equilibrium is disturbed: they can be varied, or redefined, without harming the constitutive relations. Different definitions of the hydrodynamic variables are often referred to as choosing a hydrodynamic frame~\cite{A4-Kovtun}. Traditionally, one exhausts the frame choice freedom before proceeding with any further calculation. Two common frame choices are Landau-Lifshitz~\cite{Landau} and Eckart~\cite{Eckart} frames. However, it has been known for years that first-order RH in both of these frames gives rise to the propagation of superluminal and unstable fluctuations~\cite{C1-Hiscock:1983zz, C2-Hiscock:1985zz, C3-Hiscock:1987zz}. To remedy this stability and causality (SC) problem, M\"uller, Israel, and Stewart (MIS) suggested that the d.o.f.'s of RH have to be expanded to include new dynamical variables~\cite{C4-Israel:1976qq, C5-Israel:1979wp, C6-Muller:1967zp}. For example, the shear stress tensor is promoted to a dynamical variable. In the MIS framework, the dynamics of the new variables can be obtained using the second law of thermodynamics. These variables evolve such that they relax to their first-order forms in a finite time. For the MIS framework to be stable and causal, certain parameters need to satisfy nontrivial conditions~\cite{C7-Denicol:2008ha, C8-Pu:2009fj}. The existence of such conditions may have been a motivation for further inspection of the SC problem. One could ask if a formulation of the first-order hydrodynamics exists that resolves the SC problem without the introduction of new d.o.f.'s? It has been recently claimed that such a formulation is possible~\cite{D1-Kovtun:2019hdm, D2-Bemfica:2019knx}. In this approach, one starts with the most general frame (GF) and looks for the conditions on the transport coefficients that ensure stability and causality. In other words, the GF approach expands the number of parameters instead of adding new variables. 
\par
In practice, one would solve the RH equations to obtain the evolution of hydrodynamic variables in terms of spacetime coordinates. Unsurprisingly, solving RH equations for the purpose of explaining experimental data demands numerical methods. However, there are a few simple setups that give rise to an analytical solution~\cite{Bjorken-flow, Gubser:2010ze, CGHK-cyl-flow}. The simplest setup in a QGP context is the Bjorken flow, in which every quantity is a function of a single variable i.e., the proper time. The perfect Bjorken solution can be generalized to the first-order in the Landau-Lifshitz frame. Such a generalization gives rise to a first-order linear equation that has an exact solution~\cite{Baier:2006um}. However, the generalization of the Bjorken flow to the MIS framework leads to a nonlinear second-order differential equation~\cite{E2-Heller:2013fn, E3-Heller}. The nonlinearity of the MIS equation of motion has crucial implications. Unlike linear differential equations, the nonlinear equations with different initial conditions may decay to a common attractor, and therefore lose the initial information from the regulating sector~\cite{E2-Heller:2013fn, E3-Heller, E4-Heller:2016rtz, E5-Denicol:2017lxn, E6-Basar, E7-Aniceto:2015mto, E8-Heinz:2019dbd}. Another crucial point is that the late-time expansion solution is a divergent but asymptotic series. The attractor is commonly found by rewriting the equations of motions in terms of a dimensionless function of a dimensionless time. Besides computational simplicity, this function has a straightforward relationship with the anisotropy between the longitudinal and transverse pressures. The pressure anisotropy has a physical significance and provides an unambiguous measure for hydrodynamization (see Ref.~\cite{Florkowski:2017olj} and references wherein). Unsurprisingly the earlier studies of the GF approach investigated Bjorken flow~\cite{Bemfica:2017wps,Das-Massless}. To the best of our knowledge, Ref.~\cite{Bemfica:2017wps} is the first work to suggest that a stable and causal first-order hydrodynamics may exist. The authors presented the equations of motion in the Bjorken and Gubser cases as two applications and borrowed the methods from MIS/BRSSS studies to find a numerical attractor. On the other hand, the authors of Ref.~\cite{Das-Massless} assumed a special form for the transport parameters, that breaks the conformal invariance, and constant relaxation time on the MIS side, which may violate the SC criteria, to find an equivalence between the two approaches. Both references report that matching between the two frameworks in a \textit{strictly conformal theory} is impossible. The aforementioned works leave significant issues unsettled. First, Ref.~\cite{Bemfica:2017wps} does not explain why the method that worked in the MIS formalism should also work in the GF approach. In particular, what is the physical significance of the emergence of attractors in the GF approach? Is it still related to the pressure anisotropy? Second, what is the significance of the slight differences between the equations of motion in the two formalisms? Can we match the late-time expansion in the GF approach to a microscopic theory? Third, is there any relationship between the attractors' behavior and SC conditions of the GF approach? In the current work, we address these questions by a detailed analysis of the equations and find some subtle results.
\par    
The organization of this paper is as follows: In Sec.~\ref{sec:general}, we set up the stage by a quick review of concepts in the GF and MIS formalisms. We compare the modes that appear in the linear treatment of the perturbations around a hydrostatic equilibrium. In particular, we argue that for a highly symmetric flow such as Bjorken and Gubser, the shear channel of the GF approach may be vulnerable to instabilities in a boosted frame. The section closes with the introduction of a hydrodynamization measure for the conformal flows in the GF approach. In contrast to the MIS formalism, this measure is unrelated to the pressure anisotropy. In Sec.~\ref{sec:bjorken} we solve Bjorken equations of motion in the GF approach for conformal uncharged fluids. We start our investigation by finding the gradient expansion for the temperature at late times, which resembles the one obtained using the MIS and BRSSS approaches~\cite{A1-Romatschke, Florkowski:2017olj}. It is a divergent and asymptotic expansion, which is truncated if we eliminate the off-shell\footnote{We follow the terminology of Ref.~\cite{D1-Kovtun:2019hdm}. An on-shell quantity is evaluated on the solutions to the equations of motion, and an off-shell quantity is not.} regulating sector. However, unlike the MIS/BRSSS results, the gradient expansion can only be matched up to first order with the $\NFSYM$ theory~\cite{Heller:2007, Heller:2008}. The first-order result is $\eta/s=1/(4\pi)$ as anticipated, but matching the second-order coefficients violates the SC conditions of the GF approach. This suggests that the off-shell transport parameters cannot be derived from a microscopic theory, and have no physical significance in a first-order theory. Subsequently, we utilize the hydrodynamization measure introduced in Sec.~\ref{sec:general} and use the slow-roll expansion to find an analytic expression for the attractor. Since this measure coincides with the function defined in~\cite{E2-Heller:2013fn}, the process is also similar. We find the first nonhydro mode and show that it is properly related to the analytical structure of the Borel-transformed series. We show that the attractor solutions that violate the SC conditions of Refs.~\cite{D1-Kovtun:2019hdm, D2-Bemfica:2019knx} exhibit a reheating at early times. In contrast to the causal and stable solutions, they also start with negative longitudinal pressure. Furthermore, we compare the off-shell entropy current divergence in the SC and non-SC regimes. It is only for the SC attractors that the off-shell parameters violate the second law of thermodynamics. We argue that this observation is consistent with the ideas presented in Ref.~\cite{Gavassino:2020ubn}. Finally, by finding a convergent power series solutions in early time, we show that the attractor is a pullback/forward one~\cite{Behtash:2019qtk, Heller-Phase}. We conclude the paper in Sec.~\ref{sec:conc}.
\par
We use natural units in which $\hbar=c=1$, and the metric convention is mostly plus. All equations are written in Milne coordinates namely $x^\mu=\left(\tau,x,y,\xi\right)$, in which
$
\tau = \sqrt{t^2-z^2}$ and $\xi = \inv{2}\log\left(\frac{t+z}{t-z}\right)\,
$. The line element of the flat spacetime in this coordinate system reads $$\rd{s^2}=-\rd{\tau^2}+\rd{x^2}+\rd{y^2}+\tau^2\rd{\xi^2}\,.$$
We denote the rapidity with $\xi$ to avoid confusion.
\section{General remarks}\label{sec:general}
\setcounter{equation}{0}
\par
In this section, we briefly review the GF approach and comment on its similarities and differences compared with the MIS one. The most general form of the energy-momentum tensor of an uncharged fluid reads~\cite{D1-Kovtun:2019hdm}
\begin{equation}\label{eq:general-em-tensor}
T^{\mu\nu}=\mathcal{E}u^\mu u^\nu + \mathcal{P} \Delta^{\mu\nu}+Q^\mu u^\nu + Q^\nu u^\mu + \pi^{\mu\nu}\,.
\end{equation}
Here 
$$
\mathcal{E}=\ed+f_\mathcal{E}\qquad\mathcal{P}=p+f_\mathcal{P}\,,\qquad \Delta^{\mu\nu}=g^{\mu\nu}+u^\mu u^\nu\,,
$$ 
in which $f_\mathcal{E}$ and $f_\mathcal{P}$ are dissipative corrections to the equilibrium energy density $\epsilon$ and pressure $p$, respectively. $Q^\mu$ is a vector transverse to the four-velocity $u^\mu$ and is usually called the heat flow. $\pi^{\mu\nu}$ is the shear stress tensor that is symmetric and transverse to $u^\mu$, and in most prescriptions it is traceless. The concrete form of the dissipative corrections is different in different prescriptions of dissipative hydrodynamics.
To the best of our knowledge, these different prescriptions can be formulated in the form of~ Eq.~\eqref{eq:general-em-tensor}. Examples of such prescriptions include first-order hydrodynamics in the Landau-Lifshitz~\cite{Landau} and Eckart~\cite{Eckart} frames, BRSSS~\cite{BRSSS}, Disconzi prescription~\cite{Disconzi:2013wia}, and Heller-Janik-Spalinski-Witaszczyk(HJSW) theory~\cite{HJSW}. Hereafter, we focus on the MIS in the Landau-Lifshitz frame and the GF approach as presented in Ref.~\cite{D1-Kovtun:2019hdm}. In the Landau-Lifshitz frame and up to fist order in derivatives, one has $f_{\mathcal{E}}=0$, $f_{\mathcal{P}}=-\zeta\sexp$,\footnote{$\zeta$ is the bulk viscosity.} and $Q^\mu=0$. Thus the equilibrium energy density is an eigenvalue of the energy-momentum tensor, namely $u_\mu T^{\mu\nu}=-\epsilon u^\nu$. At first order in derivatives, the shear stress tensor is $\pi^{\mu\nu}=-\eta\sigma^{\mu\nu}$~\footnote{$\eta$ is the shear viscosity. We adopt the definition of $\sigma^{\mu\nu}$ given in Ref.~\cite{D1-Kovtun:2019hdm}, which is $\Delta^{\mu\rho}\Delta^{\nu\sigma}\left(\nabla_\rho u_\sigma+\rho{\leftrightarrow}\sigma-(2/3)g_{\rho\sigma}\sexp\right)$.}. However, disturbances of the equilibrium in this first-order theory tend to be unstable, and possibly acausal. This can be revealed by assuming a locally constant value for the hydrodynamic variables. One then assumes a small perturbation for each of the hydrodynamic variables. The perturbations are Fourier-transformed as $\delta X(t,x) \sim \delta \tilde{X} \exp(-i\omega t+i\tvec{k}\cdot\tvec{x})$ and plugged into the equations of motion, and terms up to first order in the perturbation are kept. The result is a linear system that gives rise to dispersion relations for the frequencies of the perturbations in terms of their momentum i.e., $\omega=\omega(\tvec{k})$. In a causal hydrodynamic theory, these excitations, or modes, must not propagate faster than light. On the other hand, they cannot grow with time if the equilibrium state is supposed to be stable. Since the equations for the perturbation in the first-order Landau-Lifshitz frame are parabolic, the acausality of this theory is more simply revealed using a mixed Laplace-Fourier transform, viz. $\delta X(t,x) \sim \delta \tilde{X} \exp(-\omega t+i\tvec{k}\cdot\tvec{x})$~\cite{A2-Romatschke}. A crucial point in the calculation of modes is that there is no straightforward relation between the frequencies from a comoving frame to a boosted one~\cite{D1-Kovtun:2019hdm, C8-Pu:2009fj}. For example, a comoving observer finds no nonhydro mode\footnote{Modes for which $\omega(k=0)\neq 0.$}. But an observer uniformly moving with the uniform velocity $\tvec{v}_0$ with respect to the local rest frame (LRF) observes such modes\footnote{The direction of $\tvec{v_0}$ is chosen such that $\tvec{k}\cdot\tvec{v_0}$ is nonzero.}. An example is 
\begin{equation}
\omega = i\frac{w\sqrt{1-v_0^2}}{v_0^2\eta} + \orderof{(\tvec{k}\cdot\tvec{v_0})}\,,
\end{equation}
which is apparently unstable. Here $w=\epsilon+p$ is the enthalpy density in equilibrium. The MIS formalism introduces relaxation times and promotes the bulk pressure and shear stress tensor to the dynamical variables. In its simplest form the MIS equations read~\cite{C4-Israel:1976qq,C5-Israel:1979wp,C6-Muller:1967zp}
\begin{eqnarray*}
	&&\left(\tau_{\pi}u^\alpha\partial_\alpha+1\right)\pi^{\mu\nu} = -\eta\sigma^{\mu\nu}\,,\\
	&&\left(\tau_{\Pi}u^\alpha\partial_\alpha+1\right)\Pi = -\zeta\sexp\,.
\end{eqnarray*}
Here $\tau_{\pi}$ and $\tau_{\Pi}$ are the shear and bulk relaxation times respectively, and $\Pi$ is the bulk pressure. As mentioned in Sec.~\ref{sec:intro} the transport coefficients of the MIS formalism must satisfy certain conditions to ensure stability and causality~\cite{C8-Pu:2009fj}. Any other regularization scheme for the naive dissipative hydrodynamics needs to regulate both IR and UV growing modes.
\par
The essence of the GF approach is to introduce off-shell transport parameters into the energy-momentum tensor. The off-shell parameters for an uncharged fluid appear in the dissipative correction as~\cite{D1-Kovtun:2019hdm}

\begin{eqnarray}
	f_{\mathcal{E}}&=&\epsilon_1\frac{u\cdot\partial{T}}{T}+\epsilon_2 \sexp,\nonumber\\
	f_{\mathcal{P}}&=&\pi_1\frac{u\cdot\partial{T}}{T}+\pi_2 \sexp,\nonumber\\
	Q^\mu &=& \theta\left(u^\nu\nabla_\nu{u^\mu}+\inv{T}\Delta^{\mu\nu}\partial_\nu T\right)\,. 
\end{eqnarray}
 Here $\epsilon_1$, $\epsilon_2$, $\pi_1$, $\pi_2$, and $\theta$ are the off-shell parameters, which on certain conditions~\cite{D1-Kovtun:2019hdm} ensure the causality and stability of the theory.
The corrections must decay to their on-shell values if the system hydrodynamizes. As an example, $\mathcal{E}$ is not an eigenvalue of the energy-momentum tensor in the GF, since 
$
u_\mu T^{\mu\nu} = -\mathcal{E}u^\nu - Q^\nu\,.
$
This statement is also true in the Eckart frame. However, there is a sharp contrast between the GF and Eckart's frame. In the latter, the heat flow is on-shell, and the eigenvalue equation is not recovered by hydrodynamization. On the other hand, a measure of hydrodynamization in the GF approach could be the recovery of the Landau eigenvalue equation. Conformal flows with three spacelike symmetries, such as Bjorken~\cite{Bjorken-flow} and Gubser~\cite{Gubser:2010ze} ones, are exceptions to this statement. The heat flow cannot exist in such highly symmetric flows, and therefore the off-shell energy density is an eigenvalue of the energy-momentum tensor. One could argue that the similarity between the two approaches i.e., GF and MIS in such flows are at some level related to this fact, and that it has nothing to do with the boost invariance \emph{per se}. Despite this similarity, the ways to measure the hydrodynamization in the GF and MIS approaches are conceptually different. In MIS formalism, one looks for the decay of the pressure anisotropy~\cite{E3-Heller, Florkowski:2017olj}. If this quantity happens to quickly decay to the same solution despite different initial conditions, then hydrodynamization has occurred. On the other hand, there is no role for the regulating sector in the pressure anisotropy in the GF. One instead should look for the suppression of off-shell corrections. In the case of conformal Bjorken flow, this can be done by utilizing the same method used in the MIS approach~\cite{Bemfica:2017wps}. However, this may not occur in solutions with lower degrees of symmetry. 
\\
The GF transport parameters appear in two classes. One of them i.e., $\theta$, which appears in $Q^\mu$ regulates the modes in the shear channel. As $Q^\mu$ vanishes in highly symmetric flows, $\theta$ disappears from the shear channel. We emphasize that one must not assume $\theta$ is zero to eliminate $Q^\mu$. To limit our discussion to the case of conformal theories, we assume the following relations dictated by conformal invariance~\cite{D1-Kovtun:2019hdm}:

\begin{align}
&\epsilon = 3 p\,,\qquad \epsilon_i = 3\pi_i\,,\nonumber\\&\pi_2 = \pi_1/3\,,\quad p = \u{p}T^4\,\qquad \eta = \ceta s.
\end{align}

For uncharged fluids, the parameters $\u{p}$ and $\ceta$ are dimensionless constants. The conformal invariance also constrains the off-shell parameters to be proportional to $T^3$~\cite{D1-Kovtun:2019hdm}. Therefore they take the following forms:
\begin{equation}
\pi_1 = 16\u{p}\ceta\cpi T^3,\qquad \theta=16\u{p}\ceta\ctheta T^3\,,
\end{equation}
in which $\cpi$ and $\ctheta$ are pure numbers. The energy-momentum tensor of Eq.~\eqref{eq:general-em-tensor} for an uncharged conformal fluid in the GF is then translated into~\cite{D1-Kovtun:2019hdm}

\begin{widetext}
	\begin{eqnarray}
	T^{\mu\nu} 
	&=& \u{p}T^{3}\left[T + 16 \ceta\cpi\left(\frac{u^\nu \partial_\nu T}{T}+\frac{\sexp}{3} \right) \right] \left( g^{\mu\nu} + 4\, u^\mu u^\nu \right)  
	\nonumber \\&& 
	+ 16\u{p}\ceta\ctheta T^3\left[\left( u^\nu \nabla_\nu u^\mu + \frac{\Delta^{\mu\lambda}\partial_\lambda T}{T}\right)u^\nu + (\mu{\leftrightarrow}\nu)\right] 
	- 4\u{p}\ceta T^{3} \sigma^{\mu\nu} + O(\nabla^2)\,.
	\end{eqnarray}
\end{widetext}

In our parametrization, the stability and causality conditions of Ref.~\cite{D1-Kovtun:2019hdm} are translated into $\cpi>1$ and $\ctheta>1$. Now the mode of Eq.~(4.5) of Ref.~\cite{D1-Kovtun:2019hdm} in a boosted frame in the $x$-direction reads
\begin{equation}
\omega = \frac{iT\sqrt{1-v_0^2}}{v_0^2\ceta}+\orderof\left(\tvec{k}\cdot\tvec{v_0}\right)\,.
\end{equation}
The appearance of this unstable mode in the boosted frame is at least alarming. In the LRF, the two modes of Ref.~\cite{D1-Kovtun:2019hdm} in the shear channel reduce to one mode given below:
\begin{equation}
\omega = -i\frac{\ceta}{T}k^2\,.
\end{equation}
The above relation is exact and equivalent to the leading term in the shear channel in BRSSS Eq.~(3.28) in~\cite{BRSSS}. Higher-order terms only appear in the boosted frame and do not hold any content from the off-shell corrections. The regulator of the shear channel in the GF is $\theta$, while it is the relaxation time in MIS/BRSSS formalism. The only remaining off-shell parameter $\pi_1$ regulates the nonhydro mode that appears in the sound channel. This mode in the LRF reads 
\begin{equation}\label{eq:nh-sound-mode}
\omega = -i\frac{sT}{\pi_1}\,,
\end{equation}
and is exact. Higher-order terms only appear in the boosted frame and give rise to the condition $\pi_1>4\eta$~\cite{D1-Kovtun:2019hdm}. This is not the case in the MIS/BRSSS, in which higher-order terms exist in the LRF and are used to find the causality criteria~\cite{C8-Pu:2009fj, Florkowski:2017olj}
\begin{equation}\label{eq:MIS-crit}
T\tau_\pi > 2\eta/s\,.
\end{equation}
Comparing the mode in Eq.~\eqref{eq:nh-sound-mode} with a similar one in the MIS formalism $\omega=-i/\tau_{\pi}+\cdots$~\cite{Florkowski:2017olj}, we may introduce a parameter that resembles the relaxation time in the MIS formalism 
\begin{equation}
\tau_G = \frac{\pi_1}{Ts}\,.
\end{equation}
Then the criteria $\pi_1>4\eta$~\cite{D1-Kovtun:2019hdm} becomes
\begin{equation}
T\tau_G > 4\eta/s\,.
\end{equation}
One should be cautious with the above definition: It \textit{does not} mean that $\tau_G = 2 \tau_{\pi}$.
\par
At this point, let us consider the function
\begin{equation}\label{eq:gf-measure}
f=1+\frac{\dot{T}}{T\sexp}\,.
\end{equation}
We recognize that

\begin{equation}
f_\mathcal{P} = \left(16\u{p}\ceta\cpi\right)\left(T^{3}\sexp\right)\left[f-\frac{2}{3}\right]\,.
\end{equation}
Since $f_\mathcal{P}$ is proportional to the off-shell parameter $\cpi$, it must vanish if the system is hydrodynamized. This is only possible if $f$ tends to its nondissipative value, which is $2/3$. We conclude that $f$ is a good measure of hydrodynamization in the GF approach for highly symmetric conformal flows. As will be shown in Sec.~\ref{sec:attractor}, this definition coincides with a similar measure in the MIS formalism for the Bjorken flow. Also, for more complicated flows in which a closed form for the proper time is not known, we may identify the late-time limit with small values of $\sexp/\Lambda$. Here $\Lambda$ is a scale with the dimension of mass.
\\
We close this section with a final comment on the differences between the MIS and GF approaches. In the GF approach, the same off-shell parameter regulates nonhydro modes in the bulk and shear sectors. To clarify, by taking $\Pi=f_{\mathcal{P}}$ and neglecting the shear sector, we may write
\begin{equation}
\Pi - \tau_G \left(w\frac{\dot{T}}{T}+c_s^2w\sexp\right) + \zeta\sexp = 0\,.
\end{equation}     
This is not necessarily the case in MIS. In contrast, assuming that the bulk and shear relaxation times are equal is unrealistic, since they should follow different transport coefficients. However, it is not surprising if one observes a correspondence between the two frameworks by such an assumption~\cite{Das-Massive}. Also, assuming that the relaxation time in MIS is a constant~\cite{Das-Massless, Das-Massive} gives rise to a high degree of equivalence between the two frameworks. However, in such a case the causality criteria on the MIS side i.e., Eq.~\eqref{eq:MIS-crit} need $\eta/s$ to scale with temperature. 
\\

\section{Conformal Bjorken flow}\label{sec:bjorken}
\setcounter{equation}{0}
\par    
In this section, we present the late-time gradient expansion in the Bjorken flow for a conformal fluid in the GF. We show that the GF approach gives rise to a similar behavior known from the MIS framework~\cite{E3-Heller} without relying on any extra equation. However, this does not mean that the two frameworks exactly match in the frame-dependent transport coefficients. Subsequently, we investigate the attractor using the measure introduced in Eq.~\eqref{eq:gf-measure}. Also, we find a relation between the stability conditions found in Refs.~\cite{D1-Kovtun:2019hdm, D2-Bemfica:2019knx} and the early-time behavior of the solutions.
\par
As mentioned before, Bjorken symmetries~\cite{Bjorken-flow, Gubser:2010ze} eliminate the heat flow and reduce the equation of motion $\cd{\mu}{T^{\mu\nu}}=0$ to a single ordinary differential equation which reads
\begin{eqnarray}\label{eq:bjorken-eom}
&&\hspace{-1cm}4\cpi\ceta\ddot{T}+\dot{T}\left[T+4\cpi\ceta\left(\frac{2\dot{T}}{T}+\frac{7}{3\tau}\right)\right]\nonumber\\&&+\frac{T}{3\tau}\left(T+\frac{4(\cpi-1)\ceta}{3\tau}\right)=0\,.
\end{eqnarray}
Here $\dot{T}=\deriv{T}{\tau}$ and $\ddot{T}=\derivn[2]{T}{\tau}$. If we compare Eq.~\eqref{eq:bjorken-eom} with Eq.~(11) of Ref.~\cite{E6-Basar}, we observe that they are quite similar. However, there are terms in the two equations that cannot be matched. The GF transport coefficient $\cpi$ plays the role of both $\tau_{\Pi}$ and $\lambda_1$~\cite{E2-Heller:2013fn,E3-Heller,E4-Heller:2016rtz,E7-Aniceto:2015mto} in the MIS framework. 
\subsection{Hydrodynamic expansion}\label{sec:gradexp}
\par
Let us begin our investigation with the ordinary hydrodynamic expansion at late times. By inspecting Eq.~\eqref{eq:hydro-eom} at very late times, we arrive at the following ansatz for the late-time expansion of the solution~\cite{Florkowski:2017olj}:
\begin{equation}\label{eq:hydro-exp-ansatz}
T= \Lambda\tempr(\rho)\,,\qquad \rho\equiv\left(\frac{1}{\Lambda\tau}\right)^{1/3}\,,
\end{equation}
in which $\Lambda$ is a dimensionful constant. If we assume the ideal Bjorken solution then $\Lambda=\sqrt{T_0^{3}\tau_0}$. This indicates that $\Lambda$ contains the parts of initial conditions that cannot be lost at late times~\cite{Florkowski:2017olj}. Plugging the above formulation into Eq.~\eqref{eq:bjorken-eom} gives rise to 
\begin{widetext}
	\begin{equation}\label{eq:hydro-eom}
	\tempr'(\rho)-\frac{\tempr(\rho)}{\rho}-\frac{4\ceta(\cpi-1)}{3}\rho^2=\frac{4\cpi \ceta \rho^4}{3\tempr(\rho)}\left[\tempr''(\rho)-\tempr'(\rho)\left(\frac{3}{\rho}-\frac{2\tempr'(\rho)}{\tempr(\rho)}\right)\right]\,.
	\end{equation}
\end{widetext}
\par
One should bear in mind that small $\rho$ is equivalent to the late times and vice versa. As Eq.~\eqref{eq:hydro-eom} suggests, $\pi_1$ (or equivalently $\cpi$) is the source of nonlinearity in the equation of motion. This equation reduces to the linear equation of motion in the Landau-Lifshitz frame~\cite{Baier:2006um} with $\cpi=0$. Before presenting a solution to Eq.~\eqref{eq:hydro-eom}, there are some points worth mentioning: 
\begin{enumerate}[label=\Roman*.,leftmargin=0.5cm]
	\item
	The canonical entropy reads~\cite{D1-Kovtun:2019hdm}
	\begin{eqnarray}\label{eq:scanon}
	s^\mu &=& \frac{4\u{p}\tempr(\rho)^3}{\tau_0^3}\Big(1+\frac{4\cpi\ceta\rho^3}{\tempr(\rho)} \left(1-\rho\frac{\tempr'(\rho)}{\tempr(\rho)}\right),\tvec{0}\Big)
	\nonumber\\&=& s(\rho)\Big(1+\frac{4\cpi\ceta\rho^3}{\tempr(\rho)} \left(1-\rho\frac{\tempr'(\rho)}{\tempr(\rho)}\right),\tvec{0}\Big)
	\,,
	\end{eqnarray}
	which gives rise to the following relation for final produced entropy~\cite{Gubser:2010ze}:
	\begin{eqnarray}\label{eq:dsdy}
	\deriv{S}{\eta}&=&\frac{4\u{p}\pi R^2\tempr(\rho)^3}{
		\tau_0^2}{\times}\\\nonumber&&\Bigg(\inv{\rho^3}
	+\frac{4\cpi\ceta}{\tempr(\rho)} \left(1-\rho\frac{\tempr'(\rho)}{\tempr(\rho)}\right)\Bigg)\,.
	\end{eqnarray}
	In the Landau-Lifshitz frame, the $\cpi$-dependent term in Eq.~\eqref{eq:scanon} does not appear, and the entropy current of the uncharged fluid has a similar form to the perfect fluid's case, i.e., $su^\mu$. By Eq.~\eqref{eq:dsdy}, one is tempted to conclude that the distribution of the final particles holds some information from the regulating sector. However, the off-shell parameters violate the second law of thermodynamics~\cite{D1-Kovtun:2019hdm}, and therefore the regulating sector must not have any trace in the physical observables such as $dN/dy$. This is not necessarily the case for the second-order theories (see, for example, Sec.~6.6 of Ref.~\cite{Florkowski:2017olj}).
	Taken Eq.~\eqref{eq:dsdy} on shell, the extra terms from $\cpi$ vanish, and the Landau-Lifshitz frame results are recovered. This is a confirmation of the idea that physics is not frame dependent.
	\item The pressure anisotropy is given by~\cite{Florkowski:2017olj}
	$$
	\mathcal{A}=\frac{P_T - P_L}{p}\,,
	$$
	in which $P_T=T^x_{~~x}=T^y_{~~y}$, $P_L=T^\xi_{~~\xi}$ and $p$ is the equilibrium pressure.
	In the MIS formalism, the regulating sector contributes to the pressure anisotropy. By this virtue, it can be used as a measure for hydrodynamization~\cite{Florkowski:2017olj}. This is not the case in the GF formalism, in which the pressure anisotropy receives no contribution from the regulating sector
	\begin{equation}
	\mathcal{A}_{\text{GF}} = \frac{8\ceta}{T\tau}
	\end{equation}
	\item
	Equation~\eqref{eq:bjorken-eom} has an exact solution that is independent of any initial or boundary condition:
	\begin{equation}
	\tempr_{\text{exact}}(\rho)=\frac{2\ceta}{3}(16\cpi-1)\rho^3\,.
	\end{equation} 
	The occurrence of such solutions is an aspect of nonlinear differential equations and thus cannot appear in the Landau-Lifshitz frame. However, a solution that is independent of any initial or boundary conditions may not be physical. In our case, the unphysical nature of the exact solution is confirmed by the observation that it gives rise to a negative temperature in the on-shell limit.
\end{enumerate}

\begin{figure*}[hbt]\centering    
	\includegraphics[height=5cm]{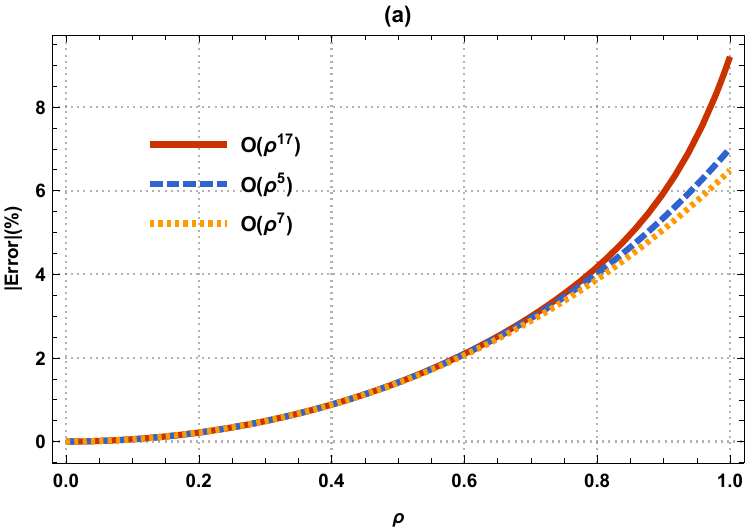}    
	\hspace{0.1cm}        
	\includegraphics[height=5cm]{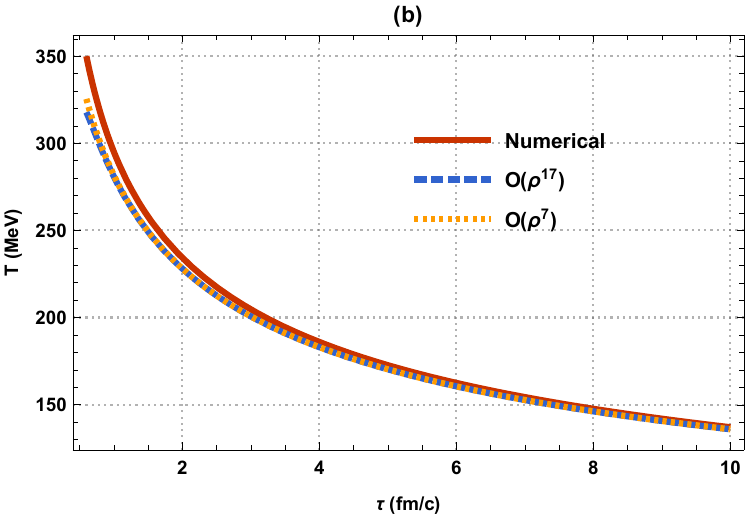}
	\caption{A numerical inspection of the hydrodynamic expansion [Eq.~\eqref{eq:small-rho-power}] with parameters given in Eq.~\eqref{eq:base-num-params}. (a) The error for different truncation, and (b) the temperature evolution. The asymptotic nature of the hydrodynamic expansion is manifest.}\label{fig:powerSeriesComp}
\end{figure*}
To find the hydrodynamic expansion, we utilize a power series around the ideal Bjorken flow and plug it into Eq.~\eqref{eq:hydro-eom}
\begin{eqnarray}\label{eq:small-rho-power}
\tempr_{N}(\rho)=C\rho+\frac{\ceta\rho^3}{3}\sum_{n=0}^{N}P_{n}(\cpi)\left(\frac{\ceta\rho^2}{3}\right)^{n}\,,
\end{eqnarray}
in which $C$ is a normalization constant and the coefficient of each power in Eq.~\eqref{eq:small-rho-power} is a polynomial in $\cpi$:
\begin{equation*}
P_{n}(\cpi)=\sum_{m=0}^{n}c_m\cpi^{m}\,.
\end{equation*}
The leading power of $\cpi$ in $P_n$ has the following form:
\begin{equation}\label{eq:lo-alpha}
c_n=-\frac{2}{C^n}8^n(n+1)!\,.
\end{equation}
The above relation shows that Eq.~\eqref{eq:small-rho-power} is a divergent but asymptotic series in $\rho\to 0$ with the optimal truncation~\cite{Bender} for $\rho \gg 1$ roughly occurring around
$
N < \frac{3C}{8\ceta}
$.
The factorial form that appears in Eq.~\eqref{eq:lo-alpha} also suggests that the series in Eq.~\eqref{eq:small-rho-power} is Borel-resumable~\cite{Bender,E6-Basar}. A careful Borel resummation of the aforementioned series requires a thorough inspection of the analytical structure of the Borel transform~\cite{E3-Heller}. Such an investigation is not discussed in the present work, but we mention that the first real pole of the Borel-transformed series occurs at $\frac{3\mathcal{C}}{8\cpi\ceta}$. When compared with the MIS results~\cite{E3-Heller,E7-Aniceto:2015mto,E6-Basar}, this trend of poles is consistent with the interpretation of $\tau_G=4\ceta\cpi/T$ as a relaxation time. That being said, a naive resummation using Eq.~\eqref{eq:lo-alpha} gives rise to a very good approximation of the solution at very late times:
\begin{eqnarray*}
	\tempr_{\text{Borel}}(\rho)&=&C\rho+\frac{C\rho}{4\cpi}\exp\left(-\frac{3C}{8\cpi\ceta\rho^2}\right)\\&&\hspace{1cm}\times\Gamma\left[0,-\frac{3C}{8\cpi\ceta\rho^2}\right]\,.
\end{eqnarray*}
The above resummation has a constant imaginary part which is anticipated by the non-sign-alternating form in Eq.~\eqref{eq:lo-alpha}. The appearance of an imaginary part shows that the resummed solution to Eq.~\eqref{eq:hydro-eom} should be obtained using a trans-series~\cite{E3-Heller}. For $C=1$, the optimal truncation occurs at $N=2$---i.e., up to $\orderof{(\rho^7)}$---which reads
	\begin{eqnarray}\label{eq:hydro-exp}
	T(\tau)&=&\Lambda\left(\frac{1}{\Lambda\tau}\right)^{1/3}\Bigg[1-\frac{2\ceta}{3}\left(\frac{1}{\Lambda\tau}\right)^{2/3}-\frac{16\cpi\ceta^2}{9}\left(\frac{1}{\Lambda\tau}\right)^{4/3}\nonumber\\&&-\frac{256\cpi^2\ceta^3}{27}\left(\frac{1}{\Lambda\tau}\right)^{2}\Bigg]\,,
	\end{eqnarray} 
\subsection{Matching with $\NFSYM$ theory}\label{sec:matching}
Equation.~\eqref{eq:hydro-exp} resembles the one that is found using the MIS approach~\cite{A1-Romatschke,Florkowski:2017olj}. At the risk of repeating ourselves, we emphasize that a match between $\cpi$ and MIS transport coefficients cannot be found by simple comparison. Furthermore, a matching between this gradient expansion and a microscopic theory fails at the second order. To clarify, let us consider the $\NFSYM$ theory and try to match the coefficients with the gravity dual~\cite{Bhattacharyya:2008jc}. By matching the second term in the bracket in Eq.~\eqref{eq:hydro-exp}, we find $\ceta=1/4\pi$. This is not surprising, since the shear viscosity is frame independent. On the other hand, matching the third term gives rise to 
\begin{equation}
\cpi=\frac{1-\log{2}}{4}\,,
\end{equation}
which violates the SC conditions of Refs.~\cite{D1-Kovtun:2019hdm, D2-Bemfica:2019knx}. Since we have exhausted all transport parameters in the GF formalism, matching the next-order coefficient is not even possible. Although we have shown that matching fails for a particular microscopic theory, there are good reasons to believe that this is a general result. A similar failure is also reported in Ref.~\cite{Das-Massless} with a different setup. However, one should not interpret this failure as a breakdown of the GF approach. The correct understanding is that the GF theory is a first-order one, and it does not need to match with any microscopic theory beyond its frame-independent parameters. But this might be a practical problem if one wants to run simulations using this approach. A practical estimate for $\cpi$ may be found as follows. Comparing the causality criteria in MIS and GF, one may assume that $\tau_G \sim 2 \tau_{\pi}$. By this virtue, a \textit{good value} for $\cpi$ reads 
$$
\cpi = 2 (2-\log{2}) \approx 2.6\,.
$$ 
This specific value for $\cpi$ is proposed as a benchmark for prospective simulations. However, we are using different values of $\cpi>1$ in what follows to emphasize that one cannot say what is the correct value for the off-shell parameters. A numerical investigation of the solutions is possible by assuming consistent initial conditions for $\tempr(0)$ and $\tempr'(0)$. The numerical results represented in Fig.~\ref{fig:powerSeriesComp} confirm the analysis given in the previous lines. The parameters used for the numerical results are
\begin{eqnarray}\label{eq:base-num-params}
&&\ceta = \frac{1}{4\pi}\,,\qquad\cpi=\frac{\pi}{3}\,,\\\nonumber&&\tau_0=0.6\fmc,\qquad T(\tau_0)=350\MeV.
\end{eqnarray}
\subsection{The attractor}\label{sec:attractor}
\begin{figure*}[!hbt]\centering    
	\includegraphics[height=5cm]{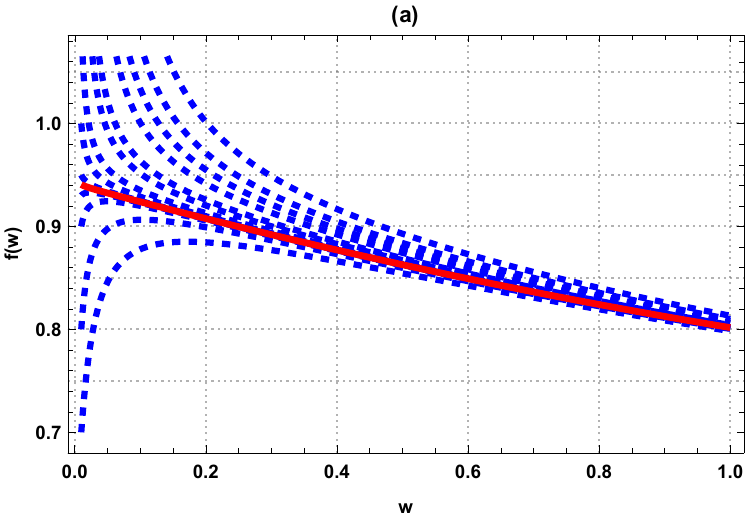}    
	\hspace{0.1cm}        
	\includegraphics[height=5cm]{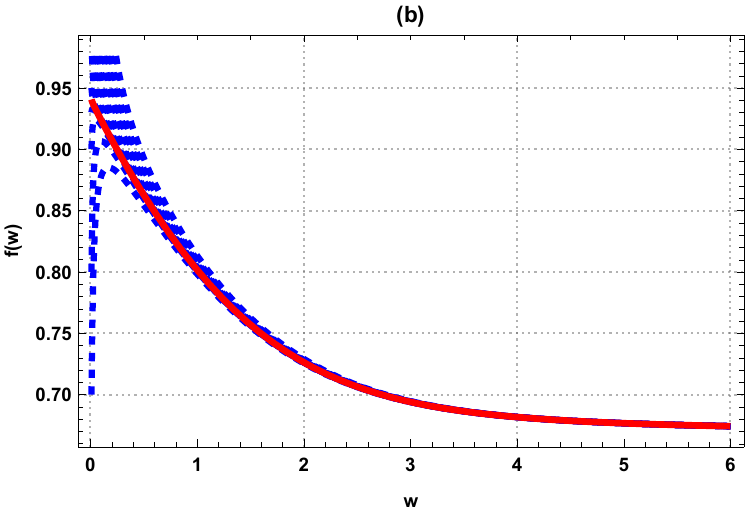}
	\caption{(a) Quick decay of solutions of Eq.~\eqref{eq:heller-eom} with different initial conditions to the attractor. The red solid line is the numerical attractor found from the initial condition given in~\eqref{eq:f0}. (b) The decay of all solutions to $f=2/3$ at late times. The transport parameters are $\cpi=2(2-\log{2})$ and $\ceta=1/4\pi$.}\label{fig:attractor}
\end{figure*}
To understand the early-time behavior of Eq.~\eqref{eq:bjorken-eom}, we work with the measure introduced in Eq.~\eqref{eq:gf-measure}. In the Bjorken case, it matches with the function introduced in Ref.~\cite{E3-Heller} and reads 
\begin{equation}\label{eq:f-def}
f(w)=1+\frac{\rd{\log{T}}}{\rd{\log{\tau}}}\,.
\end{equation}
Following Ref.~\cite{E3-Heller}, we rewrite our equations in terms of the following variables:
\begin{equation}
w=T\tau\,,\qquad f=\tau\frac{\dot{w}}{w}\,.
\end{equation}
The quantity $w$ at late times approaches to $(\Lambda\tau)^{2/3}$ and therefore tends to lose initial conditions other than the ones encoded in $\Lambda$. An interesting fact about $w$ is that it can be recognized as the Lagrangian of the perfect isentropic uncharged fluid with a minus sign i.e., $L=-w$, provided that $\tau$ is the fluid's proper time~\cite{Markakis:2016udr,Shokri:2019rsc}. Therefore, the Lichnerowicz-Carter equation of motion~\cite{Lichnerowicz,Markakis:2016udr,A3-Rezzolla} can be written as\footnote{Here $\lieder{}{}$ is the Lie derivative, $\dform{u}$ is the velocity one-form, and $\dext(w)$ is the exterior derivative of $w$.} 
$$
\lieder{\tvec{u}}{(w\dform{u})} = -\dext(w)\,.
$$
Using
$$
\dot{w}=\frac{f(w)w}{\tau},\qquad\ddot{w}=\frac{w}{\tau^2}\left(wf'(w)f(w)+f(w)^2-f(w)\right)\,,
$$
Equation~\eqref{eq:bjorken-eom} is transformed into the following first-order ODE:
\begin{eqnarray}\label{eq:heller-eom}
&&\hspace{-1.5cm}4\cpi\ceta wff'+12\cpi\ceta f^2-\frac{56}{3}\cpi\ceta f\nonumber\\&&+wf+\frac{64\cpi\ceta}{9}-\frac{4\ceta}{9}-\frac{2w}{3}=0.
\end{eqnarray}
Here $f'(w)$ denotes a derivative with respect to $w$. Although this equation is quite similar to Eq.~(14) of Ref.~\cite{E6-Basar}, no transformation can exactly reproduce the latter. This fact introduces differences in the attractor and its analytical structure.
\par
To gain insight into the behavior of the solutions of Eq.~\eqref{eq:heller-eom}, we examine the equation around $w=0$. Such an examination gives rise to\footnote{There are two solutions for $f(0)$. However, as in Ref.~\cite{E3-Heller}, only the presented one gives rise to an attractor.}
\begin{equation}\label{eq:f0}
f(w)=\frac{7\cpi+\sqrt{\cpi(3+\cpi)}}{9\cpi}+\orderof{(w)}.
\end{equation} 
To find the attractor numerically, we assume the initial condition from the above equation for $f$ at $w=\delta_0$ at a fixed value of $\cpi$ and solve Eq.~\eqref{eq:heller-eom}. Here $\delta_0$ is an arbitrary small number that can be numerically processed. Then we look if solutions with arbitrary initial conditions decay to this solution. An illustration of such a computation is represented in Fig.~\ref{fig:attractor}. As this figure suggests, the solutions with different initial conditions decay to an attractor at late times. Equation~\eqref{eq:f0} also gives rise to a relation between the stability condition for $\cpi$ and the early-time behavior of the attractors: $f(0)$ is equal to $1$ for $\cpi=1$ and blows up as $\cpi$ tends to zero. On the other hand, it has a finite lower bound when $\cpi$ tends to infinity i.e.,
$
f(0) > \frac{8}{9}.
$
\subsubsection{SC conditions, reheating, and the sign of longitudinal pressure}\label{sec:reheat}
\begin{figure*}[hbt]\centering    
	\includegraphics[height=5cm]{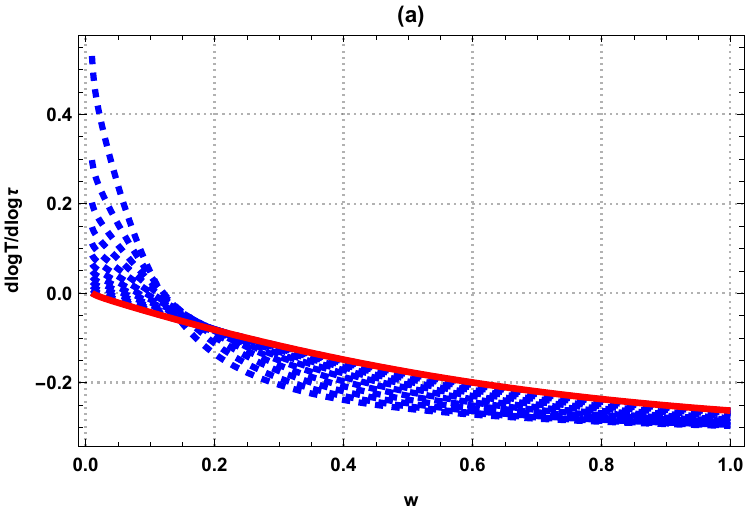}    
	\hspace{0.1cm}        
	\includegraphics[height=5cm]{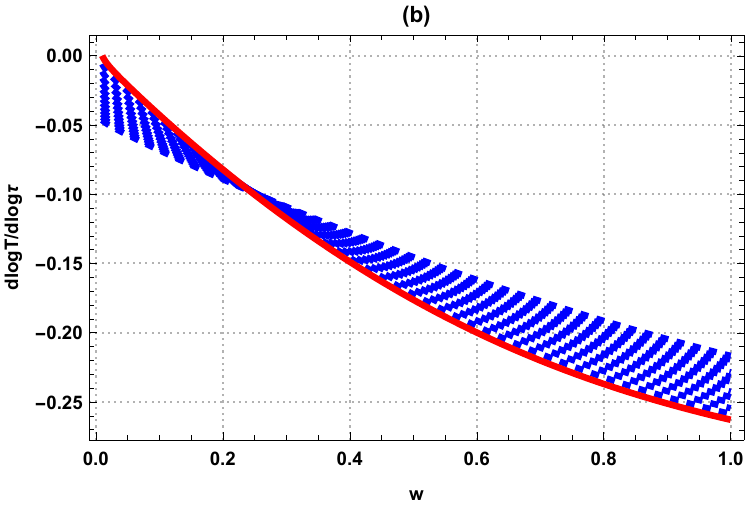}
	\caption{The depiction of $\rd{\log{T}}/\rd{\log{\tau}}$ for the attractors with (a) $\cpi<1$ and (b) $\cpi>1$. The acausal choices of $\cpi$ exhibit a reheating at early times, while this does not happen for causal values. The red solid line corresponds to $\cpi=1$.}\label{fig:reheating}
\end{figure*}

\begin{figure*}[hbt]\centering    
	\includegraphics[height=5cm]{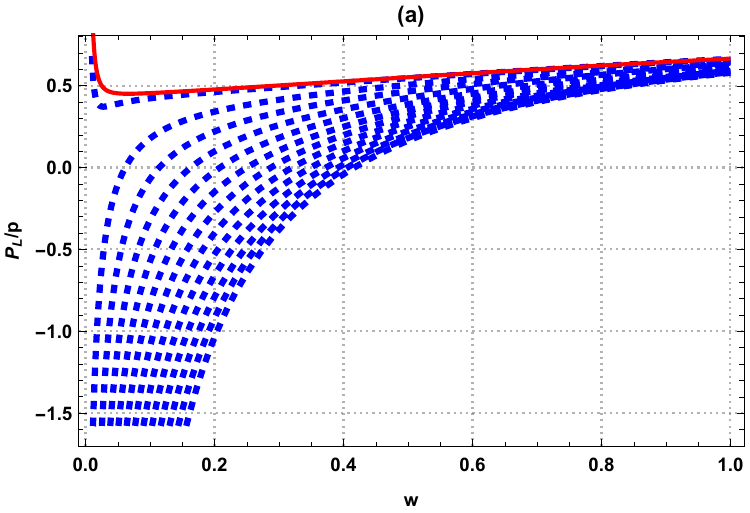}    
	\hspace{0.1cm}        
	\includegraphics[height=5cm]{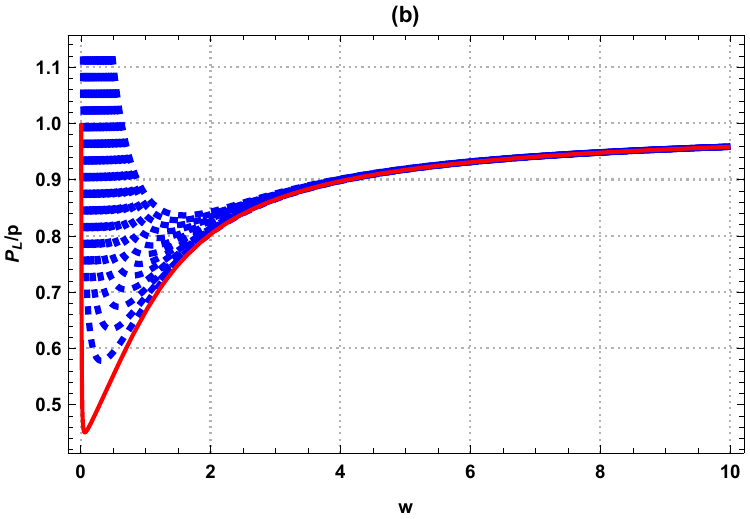}
	\caption{The depiction of $P_L/p$ for the attractors with (a) $\cpi<1$, and (b) $\cpi>1$. The acausal choices of $\cpi$ exhibit a negative $P_L$ at early times and then change sign. For causal values, this quantity is always positive. The red solid line corresponds to $\cpi=1$.}\label{fig:pl}
\end{figure*}

The values of $\cpi$ that violate the SC conditions exhibit an increase in the energy density at early times. The reheating does not happen for causal choices of $\cpi$. This is illustrated in Fig.~\ref{fig:reheating}. In the MIS formalism, the SC criteria of Eq.~\eqref{eq:MIS-crit} are not related to the existence or absence of reheating. Instead, attractors that violate Eq.~\eqref{eq:MIS-crit} may not exhibit an early reheating if $\tau_{\pi} T > \eta/s$.
Another crucial difference between causal and acausal parameters is in the change of $P_L$ sign at early times. In terms of $f$ and $w$, $P_L$ is given by 
\begin{equation}
\frac{P_L}{p} = 1 + \frac{16\ceta}{w}\left[\cpi\left(f(w)-2\right)-1\right]\,.
\end{equation}
By inspecting $P_L$ around $w=0$ using Eq.~\eqref{eq:f0}, we find that it can be negative only if $\cpi<1$. This phenomenon is illustrated in Fig.~\ref{fig:pl}. To recognize this, one can assume $\cpi=1\pm\delta_0$, in which $\delta_0$ is a very small number, and expand $P_L/p(w=0)$ with respect to $\delta_0$. The result shows that for $\cpi<1$ ($\cpi>1$) at very early times $P_L/p$ tends to negative (positive) infinity. Interestingly, for $\cpi=1$, $P_L/p$ is finite and equal to unity. The absence of negative longitudinal pressure at early times in the SC regime seems puzzling, since it does not agree with known results from the holography~\cite{Heller:2011ju}. However, one may argue that the regulating sector in the GF is off shell and cannot be matched to a gravity dual. Therefore, it is not obliged to follow such results. As for the reheating case, there is not such a relation between the sign of longitudinal pressure at early times and Eq.~\eqref{eq:MIS-crit} in the MIS formalism. In the MIS formalism, attractor solutions that do not violate Eq.~\eqref{eq:MIS-crit} can start with negative longitudinal pressures at early times.
\subsubsection{Violation of the second law of thermodynamics and SC conditions}
\begin{figure*}[hbt]\centering    
	\includegraphics[height=5cm]{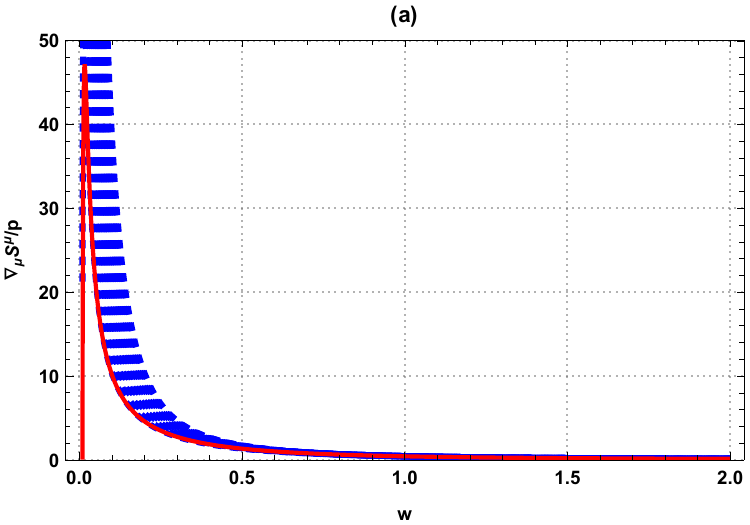}    
	\hspace{0.1cm}        
	\includegraphics[height=5cm]{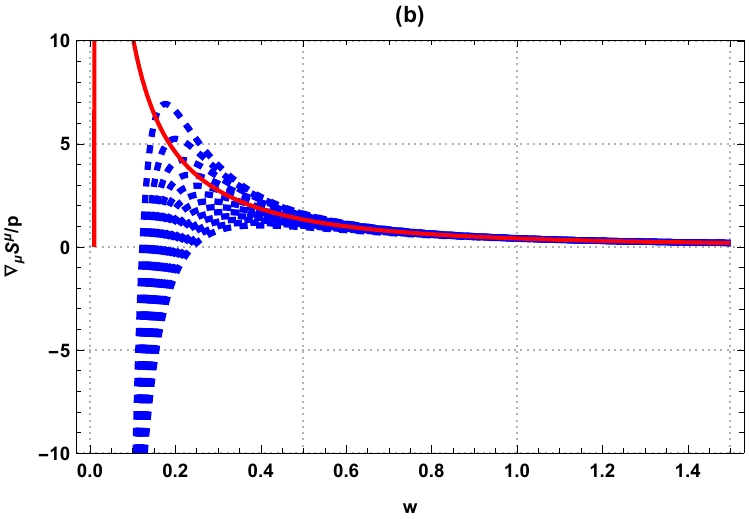}
	\caption{The depiction of $\nabla_\mu S^\mu_{can}/p$ for the attractors with (a) $\cpi<1$ and (b) $\cpi>1$. The acausal choices of $\cpi$ exhibit positive $\nabla_\mu S^\mu_{can}$ at early times and then tend to zero. For causal values, this quantity is first negative and then tends to positive values. The red solid line corresponds to $\cpi=1$.}\label{fig:divs}
\end{figure*}

In Sec.~\ref{sec:gradexp}, we mentioned that the GF off-shell parameters violate the second law of thermodynamics~\cite{D1-Kovtun:2019hdm}. This is exhibited in the early-time behavior of SC and non-SC attractors. We calculate the divergence of the canonical entropy divided by equilibrium pressure for both cases without taking the on-shell limit. For the SC attractors, this quantity is negative in early times and then tends to its on-shell positive value. On the other hand, for the SC violating attractors this quantity is always positive and boundless at early times. As in the case of longitudinal pressure in the previous section, this can be analytically realized by assuming $\cpi=1\pm\delta_0$. For small but nonzero $\delta_0$ at $w=0$, we find that in the SC (violating) regime the quantity $\nabla_\mu S^\mu_{can}$ tends to negative (positive) infinity. The transition between the two regimes discretely happens at $\cpi=1$, for which $\nabla_\mu S^\mu_{can}=0$.
This behavior is depicted in Fig.~\ref{fig:divs}. As Fig.~\ref{fig:divs} suggests the entropy current's divergence becomes positive around the dimensionless time of hydrodynamization. This confirms measuring hydrodynamization with off-shell corrections decay. Furthermore, this behavior illuminates the discussion of the canonical entropy in Ref.~\cite{D1-Kovtun:2019hdm}. As stated in Ref.~\cite{D1-Kovtun:2019hdm}, the off-shell parameters \textit{need not} satisfy the second law of thermodynamics. In light of a recent study~\cite{Gavassino:2020ubn}, we speculate that they  are \textit{required} to violate the second law to ensure the stability of the equilibrium. For the SC-violating solutions, including the Landau frame's one, there is no bound on the entropy at early times. This allows the fluctuations to run away from the equilibrium to any arbitrary state with larger entropy. However, the violation of the second law in the preequilibrium stage of the stable regime ensures that the absolute maximum entropy is reached only at equilibrium. Consequently, the fluctuations cannot destabilize it. 
\subsubsection{Slow-roll expansion and the analytical structure of the attractor}
\par
As mentioned before, $\cpi$ is the source for nonlinearity, and therefore we assume the following form~\cite{E3-Heller} to learn how the nonlinearity decays at late times:
\begin{equation}
f(w)=\frac{2}{3}+\frac{4\ceta}{9w}+\epsilon\delta f(w)\,.
\end{equation}
Plugging the above into Eq.~\eqref{eq:heller-eom} and solving it around $w\to\infty$ gives rise to
\begin{equation}
\delta f \sim \exp\left(-\frac{3w}{8\cpi\ceta}\right)w^{1+1/(4\cpi)},
\end{equation}
which can be compared with Eq.~(11) of Ref.~\cite{E3-Heller}. The exponential term can be rewritten as $\exp(-3\tau/2\tau_G)$, which is probably more comprehensive. One can also relate this term with the nonhydro mode of Eq.~\eqref{eq:nh-sound-mode}. For the above perturbation to be suppressed at late times, $\cpi$ must be positive. Putting $\cpi$ to zero and choosing the Landau-Lifshitz frame destroys the regulator that is required for the suppression of transient modes~\cite{Florkowski:2017olj}, and it is the source of the SC problem in the aforementioned frame. 
Also, coefficients in the exponential and the power must be related to the analytical structure of the Borel transformation of the asymptotic series at late times similarly to Ref.~\cite{E3-Heller}. The asymptotic series for large $w$ is equivalent to a series in small $v$ defined as
\begin{equation}
v = \frac{2}{3}\sqrt{\frac{\ceta}{w}}.
\end{equation}
Around $v=0$, the asymptotic series read
\begin{equation}\label{eq:ecalle-ltime-series}
f_N(v)=\sum_{n=0}^{N}\lambda_n v^{2n}\,,
\end{equation}
\begin{figure*}[hbt]\centering    
	\includegraphics[width=8cm]{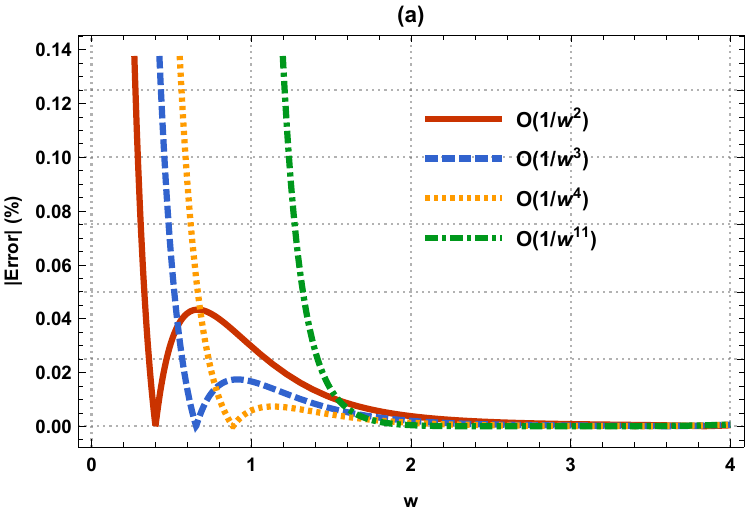}    
	\hspace{0.3cm}        
	\includegraphics[width=8cm]{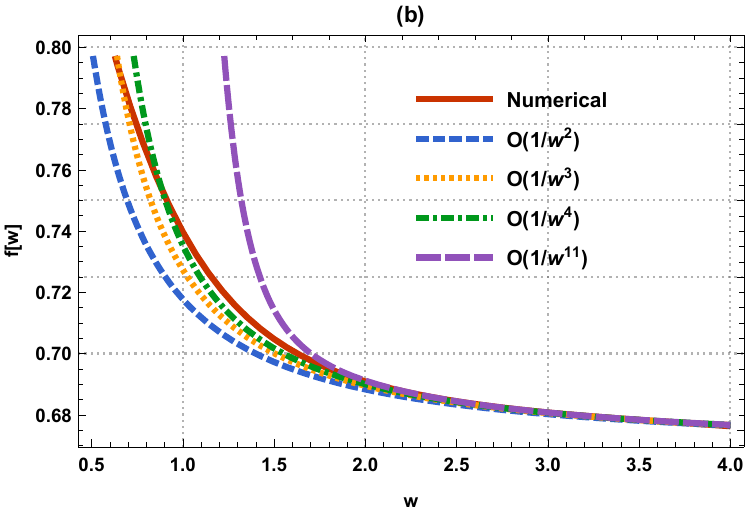}
	\caption{A numerical inspection of the late-time expansion [Eq.~\eqref{eq:ecalle-ltime-series}] with parameters given in Eq.~\eqref{eq:base-num-params}: (a) The error for different truncations. (b) $f(w)$ from numerical solution, different truncations of Eq.~\eqref{eq:ecalle-ltime-series}, and the attractor [Eq.~\eqref{eq:attractor}].}\label{fig:largeWPowerSeriesComp}
\end{figure*}
\begin{figure*}[hbt]\centering    
	\includegraphics[height=5cm]{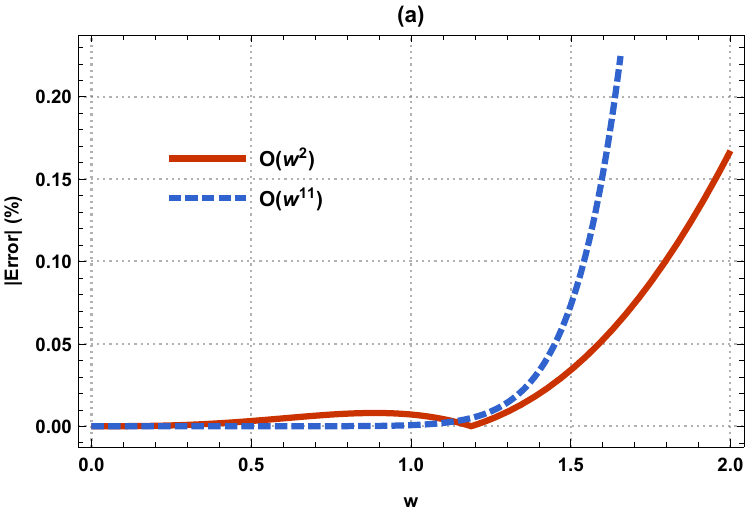}    
	\hspace{0.1cm}        
	\includegraphics[height=5cm]{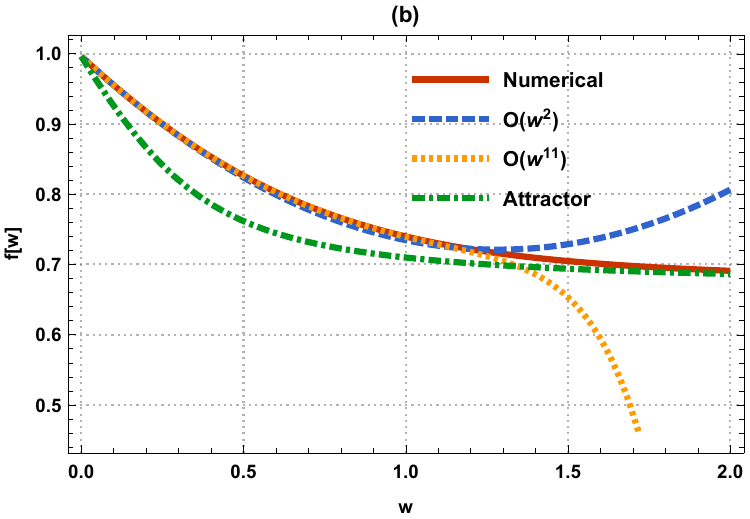}
	\caption{A numerical inspection of the early-time expansion [Eq.~\eqref{eq:ecalle-etime-series}] with parameters given in Eq.~\eqref{eq:base-num-params}: (a) The error for different truncations. (b) $f(w)$ from numerical solution, different truncations of Eq.~\eqref{eq:ecalle-ltime-series}, and the attractor [Eq.~\eqref{eq:attractor}].}\label{fig:SmallWPowerSeriesComp}
\end{figure*}
with $\lambda_0=2/3$, $\lambda_1=1$, and
\begin{eqnarray*}
	\lambda_n&=&3\cpi\Big(14\lambda_{n-1}-\frac{3(7-n)}{2}\sum_{m=0}^{n}\lambda_m\lambda_{n-m-1}\Big)\,,
\end{eqnarray*}
for $n>1$. As in Eq.~\eqref{eq:small-rho-power}, $\lambda_n$ is polynomial in $\cpi$ whose leading power is
\begin{equation}
\lambda_n=(6\cpi)^{n}(n+1)!+\cdots\,.
\end{equation}
The above form shows that the series is divergent, but asymptotic and Borel-resummable. Let the Borel-transformed series be 
\begin{equation}\label{eq:borel-f}
f_B(\xi)=\sum_{n=0}^{\infty}\frac{\lambda_n}{n!}\left[\frac{2\ceta\xi}{3}\right]^n\,.
\end{equation}
We utilize Pad\'e approximation to find the poles. Our computations up to $600$ orders in $\xi$ and with different values of $\cpi$ confirm that the first real positive pole of Eq.~\eqref{eq:borel-f} occurs at $3/(8\cpi\ceta)$. As an example, for the following set of parameters, we find the first real pole to occur at $\xi_0=1.12495$, which is in quite good agreement with $3/(8\cpi\ceta)$
\begin{equation}
\ceta = \frac{1}{4\pi}\,,\qquad \cpi=\frac{4\pi}{3}\,.
\end{equation}
We also find the power using Eq.~(17) of Ref.~\cite{E3-Heller} which for the order of the pole $-\gamma$ gives rise to $1.03312$. The latter exhibits an error of $2.5\%$ in comparison to the analytical value $1+1/(4\cpi)$. Using this information, one can reproduce the results of Refs.~\cite{E3-Heller,E6-Basar,E7-Aniceto:2015mto} for the GF approach. We find the analytical form for the attractor by assuming $\epsilon(w)/\delta= f'/f$ in which $\delta$ is an arbitrary small number. Plugging this form into Eq.~\eqref{eq:heller-eom} and expanding it in terms of $\epsilon$ in the leading order gives rise to two solutions for $f(w)$. We then expand each solution around $w\to\infty$ and compare the results with late-time expansion given in Eq.~\eqref{eq:ecalle-ltime-series}. By this comparison, we find the approximate attractor to be 
\begin{widetext}
	\begin{eqnarray}\label{eq:attractor}
	f(w)&=&\frac{2}{3}-\frac{w}{24\cpi\ceta}+\frac{8\cpi\ceta+\sqrt{192\cpi\ceta^2+(3w-8\cpi\ceta)^2}}{27\cpi\ceta}\,.
	\end{eqnarray}
\end{widetext}
A numerical representation of the late-time asymptotic series of Eq.~\eqref{eq:ecalle-ltime-series} and the approximate attractor of Eq.~\eqref{eq:attractor} for parameters given in~\eqref{eq:base-num-params} is depicted in Fig.~\ref{fig:largeWPowerSeriesComp}. The late-time expansion has a small error for $w>w_0\sim 0.5$. The optimal truncation is of order $1/w^4$, and then the approximation gets worse. The value of $w_0$ increases with $\cpi$. These comparisons confirm that Eq.~\eqref{eq:attractor} is a forward attractor~\cite{Heller-Phase}.
\par  
To check if our attractor is also a pullback one, we write the solution for early times as a series around $w=0$
\begin{equation}\label{eq:ecalle-etime-series}
f(w)=f(0)+\sum_{n=0}^{\infty}a_n\left(\frac{3w}{\cpi\ceta}\right)^n,\ 
\end{equation}
with
\begin{eqnarray}\label{eq:ecalle-etime-rec}
&&a_0=-\frac{1}{84}\,,\\\nonumber &&a_n = \frac{a_{n-1}+6(7+n)\sum_{m=0}^{n-1}\left(a_{m}a_{n-(m+1)}\right)}{56-12(7+n)f(0)}\,.
\end{eqnarray}
The numerical results for the set of parameters given in Eq.~\eqref{eq:base-num-params} are represented in Fig.~\ref{fig:SmallWPowerSeriesComp}. The early time expansion [Eq.~\eqref{eq:ecalle-etime-series}] has a marginal error up to $w_0\sim 2$, whose value increases with ${\cpi}$. The recursive relation [Eq.~\eqref{eq:ecalle-etime-rec}] and numerical computation of coefficients up to very high orders suggest that the early-time expansion has a finite radius of convergence. The good agreement between Eq.~\eqref{eq:attractor} and early-time expansion confirms that our attractor is also a pullback one.
\section{Concluding remarks}\label{sec:conc}
\setcounter{equation}{0}
\par    
In the present work, we investigated the GF approach to stable and causal first-order hydrodynamics presented in Refs.~\cite{D1-Kovtun:2019hdm, D2-Bemfica:2019knx} using Bjorken conformal flow as a toy model. We showed that this approach introduces the nonlinearity required for the hydrodynamization that is traditionally incorporated using the MIS approach~\cite{E3-Heller}. The nonlinearity seems to be a requirement for a theory that describes the transition from preequilibrium to equilibrium in a physical system. In hydrodynamics, such nonlinearity has significant consequences. First, it may give rise to a divergent asymptotic series solution at late times. In the MIS formalism, this late-time expansion can be matched to microscopic theories such as $\NFSYM$ or kinetic theory that leads to acceptable values for the relaxation time. However, as we showed in the present work, the GF's late-time expansion fails to match with $\NFSYM$ results beyond the first order. We argued that this is because the GF framework in Refs.~\cite{D1-Kovtun:2019hdm, D2-Bemfica:2019knx} is formulated up to first-order in derivatives, and its off-shell regulators do not have a physical significance~\cite{D1-Kovtun:2019hdm}. By extending the GF formalism to the next order, one can probably match the late-time expansion at higher orders. However, this may expand the number of parameters impractically.
Second, the nonlinear equations may exhibit an attractor behavior. In a simple sense, this means that their solutions lose the initial information and decay to a common solution. In stable and causal dissipative hydrodynamics, this gives rise to hydrodynamization. Hydrodynamization may be defined as the fading of initial information within the regulating sector at late times.  However, examining the attractor behavior is not a trivial task. One must find appropriate measures that are manifestly related to the hydrodynamization, and solve the equations of motions for them. As an example, the attractor behavior of the conformal Bjorken flow in MIS formalism is well understood using the pressure anisotropy as the hydrodynamization measure~\cite{Florkowski:2017olj}. We showed that pressure anisotropy cannot be used as a hydrodynamization measure in the GF approach, and we introduced a different quantity that measures the off-shell corrections decay. Fortunately, this conceptually different measure coincides with the one used in the MIS formalism for the conformal Bjorken case. Using previously introduced techniques in the MIS framework~\cite{E3-Heller, E6-Basar, E7-Aniceto:2015mto}, we presented an approximate analytical form for the attractor in the general frame approach. Although the overall behavior of the two approaches exhibits a degree of similarity, they cannot be assumed  to be equal without some extra assumptions. Such assumptions are made in Refs.~\cite{Das-Massless, Das-Massive}. 
\par
The illustrated differences between the two frameworks in a conformal theory are consistent with the frame dependency of temperature. However, the emergence of attractors in the GF approach strengthens the idea that late-time hydrodynamic behavior is independent of the chosen regulating scheme. We also showed that the SC conditions of Refs.~\cite{D1-Kovtun:2019hdm, D2-Bemfica:2019knx} are reflected in the behavior of the attractors. The SC-violating attractors exhibit reheating and negative longitudinal pressure at early times, in contrast to the stable and causal ones. Such a relation between the SC criteria and the behavior of attractors does not exist in the MIS formalism. At the moment, we cannot provide a solid physical explanation for the observed relation between SC conditions and attractors' behavior. We hope that further investigations will shed light on this.
\par
We propose some directions that go beyond the current work. The first one is to break the boost invariance to reveal the role of the heat flow regulator. The phase space of a non-boost-invariant solution is larger than the Bjorken flow. Our preliminary results show that a thorough semianalytical investigation of such flows may not be doable with the methods used in the present work. A crude examination of this question is given in the Appendix. Another notable problem is to check if the attractor behavior also happens in a nonconformal theory~\cite{Florkowski:2017olj}. One possible way is to examine a solution of Navier-Stokes equations in the Landau-Lifshitz frame, like Ref.~\cite{Csorgo:2020iug}, in the GF and/or MIS formalism. If the attractor behavior is universal, it should be observed in any prescription of the causal and stable hydrodynamics, including Disconzi's~\cite{Disconzi:2013wia}. Our spectacular examination of this prescription for the conformal Bjorken flow gives rise to results that are quite different from other approaches.  
\par
As a final comment, the second-order theories have been widely and successfully used in hydrodynamics simulations~\cite{A1-Romatschke}. It is legitimate to ask if the GF approach provides a practical alternative. In our opinion, the approach toward answering this question can start from a numerical solution of the GF equations for the Bjorken setup using a realistic equation of state. Even in such a simple setup, the breakdown of conformal symmetry introduces a large amount of arbitrariness in the choice of parameters. On the other hand, in the MIS-like theories, one can choose starting values for the second-order parameters from holography or kinetic theory~\cite{A1-Romatschke,Florkowski:2017olj}.
\begin{acknowledgments}
	M.S. thanks U.~Heinz, J.~Komijani, J.~Noronha, D.~Rischke, N.~Sadooghi, and M.~Strickland for fruitful discussions. He also thanks R.~Ryblewski for drawing his attention to their similar work~\cite{Das-Massless} and his thoughtful comments. M.S. thanks L.~Rezzolla for the hospitality during his stay at the Goethe University of Frankfurt, where the final preparation of the first version of the present manuscript is performed. The authors are particularly grateful to M.~Spalinski for discussions that overwhelmingly contributed to the revision of this manuscript. We thank the anonymous referee for their constructive comments.
\end{acknowledgments}
\appendix
\section*{APPENDIX: Comments on 1+1 self-similar flow}\label{app:A}
\setcounter{equation}{0}
\par
A minimal approach to boost invariance breaking is introduced in the 1+1 self-similar solution to ideal hydrodynamics~\cite{CGHK-cyl-flow}, which is based on the observation that although the pressure is required to be boost invariant in the Bjorken flow, the temperature and entropy density can have $\eta$ dependencies. However, it is required that their $\eta$ dependency cancel out such that the pressure remains boost invariant. Such scaling is invalid for a conformal fluid and one needs to assume an equation of state of the form 
\begin{equation}
\mathcal{E}=\kappa \mathcal{P}\,,
\end{equation} 
in which $\kappa$ is constant. To make the problem manageable, one may assume that the above equation gives rise to
\begin{equation}
\epsilon = \kappa p\,\qquad \epsilon_i = \kappa \pi_i\,.
\end{equation}
Also, it is required to assume that the temperature and entropy density have separable functionality in $\tau$ and $\eta$
\begin{equation}
T=F(\tau)\mathcal{T}(\eta),\qquad s=\frac{G(\tau)}{\mathcal{T}(\eta)}\,,\qquad p=\frac{Ts}{\kappa+1}\,,
\end{equation}
and assume some appropriate forms for the transport coefficients. The heat flow in Milne coordinates reads
\begin{equation}
Q_\mu = \theta(\tau,\eta)\frac{\mathcal{T}'(\eta)}{\mathcal{T}(\eta)}\Big(0,0,0,1\Big)\,.
\end{equation}
Plugging the above into the Euler equation gives rise to
\begin{equation}
\theta(\tau,\eta)\sim\frac{1}{\tau}\theta(\eta)\,.
\end{equation}
Assuming $\theta(\eta)=\theta_0$, the energy equation becomes separable in $\tau$ and $\eta$ with the following immediate result:
\begin{equation}\label{eq:gaussian}
\mathcal{T}(\eta)\sim\exp\left(-\frac{\eta^2}{\theta_0(1+\kappa)}\right).
\end{equation}
To obtain the above form, we have assumed that $\eta=\ceta G(\tau)$, with $\ceta$ being constant and other coefficients to be boost invariant. What remains is one equation for two unknowns, $F$ and $G$, that cannot be solved unless we assume a relation between them. From a phenomenological perspective, the Gaussian form appearing in Eq.~\eqref{eq:gaussian} looks plausible. However, in the spirit of GF, it is maybe regarded as a hint on how the perturbations decay. We believe that a comprehensive investigation of boost invariance breakdown is crucial for a thorough understanding of the GF approach.
\linebreak

\end{document}